\documentclass[pra,twocolumn,aps,showpacs,superscriptaddress]{revtex4}

\usepackage{amsmath}
\usepackage{amssymb}
\usepackage{hyperref}
\usepackage{graphicx}
\usepackage{subfigure}
\usepackage[usenames,dvipsnames]{color}
\usepackage{setspace}

\begin{document}

%%%%%%%%%%%%%%%%%%%%%%%%%%%%%%%%%%%%%%%%%%%%%%%%%%%%%%%%%%%%%%%%%%%%%%%%%%%%%%
%%%%                     Title and authors                                %%%%
%%%%%%%%%%%%%%%%%%%%%%%%%%%%%%%%%%%%%%%%%%%%%%%%%%%%%%%%%%%%%%%%%%%%%%%%%%%%%%

\title{Generation of vortex dipoles in superfluid Fermi gas in BCS limit}

\author{S. Gautam} 
\affiliation{Physical Research Laboratory,
             Navarangpura, Ahmedabad - 380 009, India}

%%%%%%%%%%%%%%%%%%%%%%%%%%%%%%%%%%%%%%%%%%%%%%%%%%%%%%%%%%%%%%%%%%%%%%%%%%%%%%
%%%%%%%%%%                    Abstract                             %%%%%%%%%%%
%%%%%%%%%%%%%%%%%%%%%%%%%%%%%%%%%%%%%%%%%%%%%%%%%%%%%%%%%%%%%%%%%%%%%%%%%%%%%%

\date{\today}
\begin{abstract}
 We theoretically investigate the generation of the vortex dipoles in superfluid
Fermi gas in the BCS limit. The vortex dipoles are generated in superfluid 
either by moving an obstacle above a critical speed or due to the decay of the 
shock waves obtained on the sudden mixing of two superfluid fragments. We 
observe that in pancake-shaped traps, the shock waves can lead to the 
formation of density ripples, which decay into vortex dipoles due to the onset 
of snake instability.
\end{abstract}


\maketitle

%%%%%%%%%%%%%%%%%%%%%%%%%%%%%%%%%%%%%%%%%%%%%%%%%%%%%%%%%%%%%%%%%%%%%%%%%%%%%%
%%%%%%%%%%%%%                 Introduction                         %%%%%%%%%%%
%%%%%%%%%%%%%%%%%%%%%%%%%%%%%%%%%%%%%%%%%%%%%%%%%%%%%%%%%%%%%%%%%%%%%%%%%%%%%%

\section{Introduction}
\label{I}
 One of the unique feature of the superfluids is that these can support only
quantized vortices. In this context, the experimental observation of the 
vortex lattice across the Bardeen-Cooper-Schrieffer to Bose-Einstein 
condensate (BCS-BEC) cross-over has unequivocally proved the superfluid nature
of these systems \cite{zwierlein2005vortices}. In the experiment, vortices 
were created by rotating the trapped superfluid Fermi gas using a blue detuned 
laser beam. In a recent experiment \cite{PhysRevLett.104.160401} with 
pancake-shaped BEC, a pair of vortices with opposite circulation were 
generated by moving an obstacle potential above a critical speed, which is 
equal to the fraction of the local sound speed. This vortex-antivortex pair is
nothing but two dimensional analogue of vortex rings. In Bose-Einstein 
condensates (BECs), vortex rings have been experimentally generated from the 
dark solitons, which undergo long wavelength transverse instability called 
snake instability \cite{anderson2001watching}. Vortices and vortex dipoles can
also be generated by merging and interfering of multiple trapped BECs 
\cite{PhysRevLett.98.110402}. The decay of shock waves into vortex dipoles has 
also been observed in BECs \cite{dutton2001observation}. The decay is mediated
by the formation of solitons which decay via snake instability. In the 
non-linear Kerr-like medium, the propagation, non-linear response, and 
collisions between the dispersive superfluid-like shock waves have been 
experimentally observed \cite{wan2006dispersive}. Shock waves have also been 
generated by merging or splitting BECs non-adiabatically 
\cite{PhysRevLett.101.170404}. The same method has been used to generate the 
shock waves in superfluid Fermi gas at unitarity \cite{joseph2011observation}.
Unlike BECs, the soliton mediated decay of shock waves into vortex dipoles has
not been unambiguously detected in superfluid Fermi gases. In a recent work 
\cite{PhysRevLett.108.150401}, formation of shock waves and domain walls in 
strongly interacting Fermi superfluid has been investigated theoretically, 
and the possibility of snake instability was conjectured in sufficiently 
wide traps. The formation and dynamics of sub- and supersonic shock waves, 
using zero temperature equations of generalized superfluid hydrodynamics, 
have also been studied in unitary Fermi gas \cite{salasnich2011supersonic}.

 Although vortex dipoles and vortex rings are yet to be experimentally 
realized in superfluid Fermi gases, it is well established that vortex
antivortex pairs play very important roles in superfluid turbulence 
\cite{barenghi2001quantized} and Berezinskii Kosterlitz Thouless (BKT) 
phase transition \cite{Berezinskii,Kosterlitz}. Recently using local 
extension of the time-dependent density functional theory, real time dynamics 
of quantized vortex rings in unitary Fermi gas in cylindrical traps was 
theoretically investigated in Ref. \cite{bulgac2011real}. 

 The advent of ultracold fermions in optical lattices 
\cite{chin2006evidence}, which are almost pure realization of Hubbard 
model \cite{hubbard1963electron}, has lead to the flurry of research
investigations. These systems are now routinely used, both by experimentalists
and theorists, to understand the strongly correlated systems. In this context,
Mott insulator of fermionic atoms in an optical lattice has already been 
experimentally realized \cite{Schneider05122008,jordens2008mott}. On the 
theoretical front, dynamical mean field theory (DMFT) \cite{RevModPhys.68.13} 
has been successfully used to study these strongly correlated systems in three 
dimensions \cite{PhysRevLett.62.324, PhysRevLett.100.056403}. The DMFT
neglects the nonlocal correlations, which can no longer be done in lower 
dimensions and frustrated systems. For these systems, cluster extension of 
DMFT like dynamical cluster approximation (DCA) and cellular dynamical 
mean-field theory (CDMFT) have been used to study these systems 
\cite{PhysRevA.82.043625,PhysRevB.82.245102,PhysRevLett.108.246402}. 
A key recent development in the field, has been the experimental realization
of spin-orbit coupling (SOC) 
\cite{PhysRevLett.109.095301,PhysRevLett.109.095302}.
The interplay between SOC and inter-atomic interactions 
leads to many interesting phenomena \cite{doi:10.1142/S0217979212300010}.
Recently, finite temperature phase diagram of two-component atomic Fermi gas
with population imbalance in the presence of Rashba spin orbit coupling
has been studied in Ref. \cite{PhysRevLett.108.080406}.

 In the present work, we numerically study the generation of vortex dipoles in 
an oblate superfluid Fermi gas (SFG) in BCS limit. We employ two methods, 
namely (a) vortex dipole generation by a moving obstacle and (b) soliton 
mediated decay of shock waves, to this end. At $T=0$K, when the normal 
component in the superfluid tends to zero, a theoretical approach based on the 
Gallilei-invariant density functional theory has been developed in 
Refs. \cite{salasnich2009hydrodynamics, Kim2004397,PhysRevA.70.033612,
PhysRevA.71.033625,PhysRevA.73.065601,springerlink:10.1134/S1054660X07020211,
adhikari2007tightly,PhysRevA.76.043626,PhysRevA.77.043609} to study the 
superfluid Fermi gases. This theoretical approach allows one to write a 
non-linear Schr\"odinger equation (NLSE) for the BCS superfluid that leads 
to the same superfluid density as the original many body fermion system 
\cite{Kim2004397}. For the applicability of this NLSE for the BCS superfluid,
the characteristic wavelength of the phenomenon under study must be larger than
the healing length. The NLSE has been used to study the collective excitations 
\cite{PhysRevA.71.033625, Kim2004397}, free expansion of the superfluid Fermi
gas \cite{PhysRevA.73.065601}, superfluid-insulator transition 
\cite{adhikari2008josephson}, and solitons 
\cite{adhikari2007tightly, PhysRevA.76.043626}.
 
 The paper is organized as follows. In Sec. \ref{II} we study the evolution
of the BCS superfluid with a repulsive obstacle moving across it. In 
Sec. \ref{III} we study the generation of the shock wave by non-adiabatically 
merging two superfluid fragments followed by the decay of shock waves into 
vortex dipoles. This is followed by conclusions in Sec. \ref{IV}.

%%%%%%%%%%%%%%%%%%%%%%%%%%%%%%%%%%%%%%%%%%%%%%%%%%%%%%%%%%%%%%%%%%%%%%%%%%%%%%
%%%%   Generation of the vortex dipoles by the repulsive potentials     %%%%%%
%%%%%%%%%%%%%%%%%%%%%%%%%%%%%%%%%%%%%%%%%%%%%%%%%%%%%%%%%%%%%%%%%%%%%%%%%%%%%%

\section{Generation of the vortex dipoles by the repulsive potentials}
\label{II}
Barring normalization factors, the order parameter for the superfluid bosonic 
and fermionic systems are
\begin{eqnarray}
\Xi_{\rm B}(\mathbf r, t) & = & \langle\hat\psi(\mathbf r, t)\rangle,
                                \nonumber\\
\Xi_{\rm F}(\mathbf r, t) & = & \langle\hat\psi_\downarrow(\mathbf r, t)
                                \hat\psi_\uparrow(\mathbf r, t)\rangle,
\end{eqnarray}
respectively, where $\psi(\mathbf r, t)$ and $\psi_\sigma(\mathbf r, t)$ with
$\sigma = \uparrow$ or $\downarrow$ are the bosonic and fermionic annihilation
field operators, respectively. In case of bosons, the order parameter is the 
wavefunction of the macroscopically occupied single particle state. Similarly,
the order parameter can be considered as the wavefunction of the 
macroscopically occupied two particle state for Fermi superfluids. The order 
parameter is normalized to total number of condensed bosons 
\cite{landau1987statistical} or condensed Cooper pairs \cite{leggett2006,
giorgini2008theory,PhysRevA.72.023621,PhysRevA.76.015601}. In the BCS limit, 
the $s$-wave scattering length between two hyperfine states of the Fermi gas 
$a\rightarrow -0$. In this limit, the superfluid Fermi gas, consisting of 
equal number of two components, at $T=0$ K is 
described by mean field equation
\cite{salasnich2009hydrodynamics, Kim2004397,PhysRevA.70.033612,
PhysRevA.71.033625,PhysRevA.73.065601,springerlink:10.1134/S1054660X07020211,
adhikari2007tightly,PhysRevA.76.043626,PhysRevA.77.043609}
\begin{eqnarray}
 \left[-\frac{\hbar^2}{4 m} \nabla^2+ \frac{2\hbar^2}{m_p} (3\pi^2)^{2/3}
 |\psi(\mathbf r,t)|^{4/3}  +  V(\mathbf r,t)\right]\psi(\mathbf r,t)
  \nonumber\\ 
  = i\hbar\frac{\partial\psi(\mathbf r,t)}{\partial t},~~~~
\label{nlse}
\end{eqnarray}
where $\psi(\mathbf r,t)$ is complex order parameter for the SFG, $m$ is the 
atomic mass of the fermionic species, $m_p=2m$ is the mass of the fermionic pair, 
and $V(\mathbf r)$ is the trapping potential. In the aforementioned equation, 
$\psi(\mathbf r,t)$ is normalized to the number of fermions $N$, i.e., 
$\int |\psi(\mathbf r,t)|^2d\mathbf r = N$; this normalization is different 
from the normalization of the $\Xi_{\rm F}(\mathbf r, t)$ 
\cite{landau1987statistical,PhysRevA.72.023621,PhysRevA.76.015601}. In terms 
of $\psi(\mathbf r,t)$ the energy of the SFG is 
\cite{salasnich2009hydrodynamics,springerlink:10.1134/S1054660X07020211}
\begin{equation}
E = \int\left[\frac{\hbar^2}{4 m} |\nabla \psi|^2+ \frac{3\hbar^2}{5m} (3\pi^2)^{2/3}
     |\psi|^{10/3}  +  V(\mathbf r,t)|\psi|^2\right]d\mathbf r,\nonumber 
\end{equation}
here $\psi = \psi(\mathbf r,t)$.
In the present work, we consider pancake-shaped trapping potential 
\begin{equation}
 V(\mathbf r,t) = \frac{m_p\omega^2}{2}(x^2+y^2+\alpha^2 z^2) 
                  + V_{\rm obs}(\mathbf r,t),
\end{equation}
where $\omega$ is the radial trapping frequency, $\alpha\gg1$ is the ratio of 
axial to radial trapping frequency, and $V_{\rm obs}$ is the obstacle 
potential which can be attractive or repulsive. We consider SFG of $^{40}$K
with $N = 10^3$, $\omega = 10$Hz, $\alpha = 10$, and
\begin{equation}
V_{\rm obs}(\mathbf r,t) = V_0 \exp\left\lbrace -2\frac{\left[(x-x_0(t))^2
                         +(y-y_0(t))^2\right]}{w_0^2}\right\rbrace,\nonumber
\end{equation}
here $\left(x_0(t),y_0(t),0\right)$ is the instantaneous location of the Gaussian 
obstacle potential with $1/e^2$ width equal to $w_0$. In Ref. 
\cite{buitrago2009mean}, effective two dimensional (2D) equations for the 
pancake-shaped Fermi superfluids in both the BCS and unitary limit 
($a\rightarrow -\infty$) have been proposed. These 2D equations provide a
good approximation to the real three dimensional (3D) systems. In the present
work, instead of using these 2D equations, we consider the full three 
dimensional equation and solve it numerically using the split time step 
Crank-Nicolson method \cite{muruganandam2009fortran}. We can rewrite the
Eq. \ref{nlse} in scaled units using the transformations
\begin{eqnarray}
\mathbf r  =  \mathbf r' a_{\rm osc},~t = t' \omega^{-1},
                                                              \nonumber\\ 
\psi(\mathbf r, t) = \frac{\sqrt{N}\phi(\mathbf r', t')}{a_{\rm osc}^{3/2}}, 
\end{eqnarray}
where the primed quantities are in scaled units and $a_{\rm osc} = 
\sqrt{\hbar/(m_p \omega)}$ is the oscillator length. After dropping the 
primes, the scaled NLSE describing the superfluid Fermi gas in BCS limit is
\begin{eqnarray}
 \left[-\frac{\nabla^2}{2} + 2 (3\pi^2 N)^{2/3}
 |\phi|^{4/3}+V(\mathbf r,t)\right]\phi(\mathbf r,t) \nonumber\\ 
 = i \frac{\partial\phi(\mathbf r,t)}{\partial t},
\label{nlse_scaled}
\end{eqnarray}

\begin{figure}[h]
 \includegraphics[width=8.5cm] {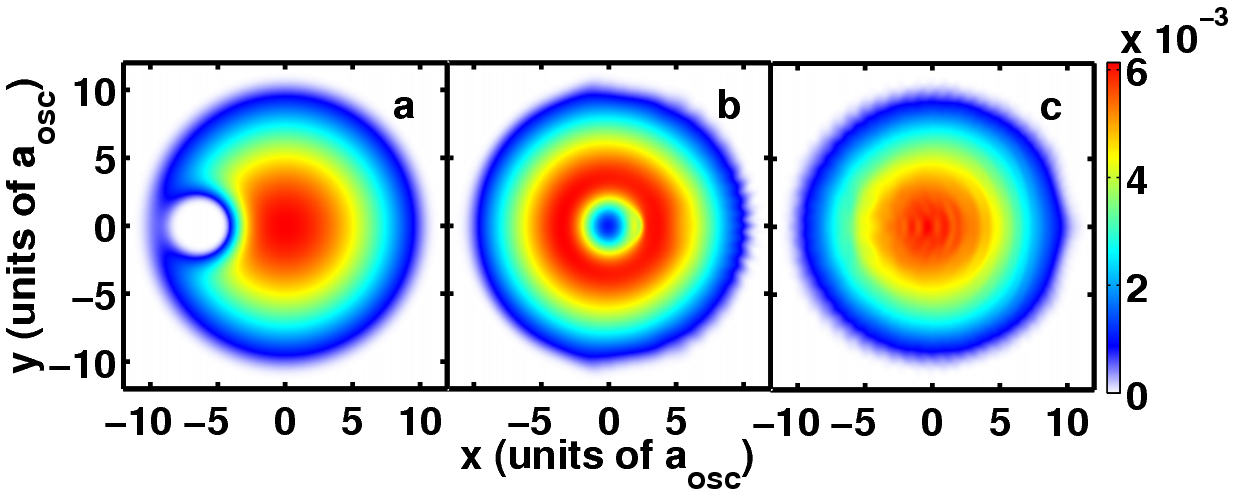}
 \caption{(Color online) Evolution of the superfluid Fermi gas in BCS limit
         when the obstacle is moving with the sub-critical speed of $150~\mu$m/s.
         Images with labels (a), (b), (c) are the superfluid densities 
         $|\phi(\mathbf r,t)|^2$ in units of $a_{\rm osc}^{-3}$ on $xy$ plane 
         at $t=0$ ms, $t=143$ ms, and $t=286$ ms respectively. 
         }
\label{fig1}
\end{figure}

\begin{figure}[h]
 \includegraphics[width=8.5cm] {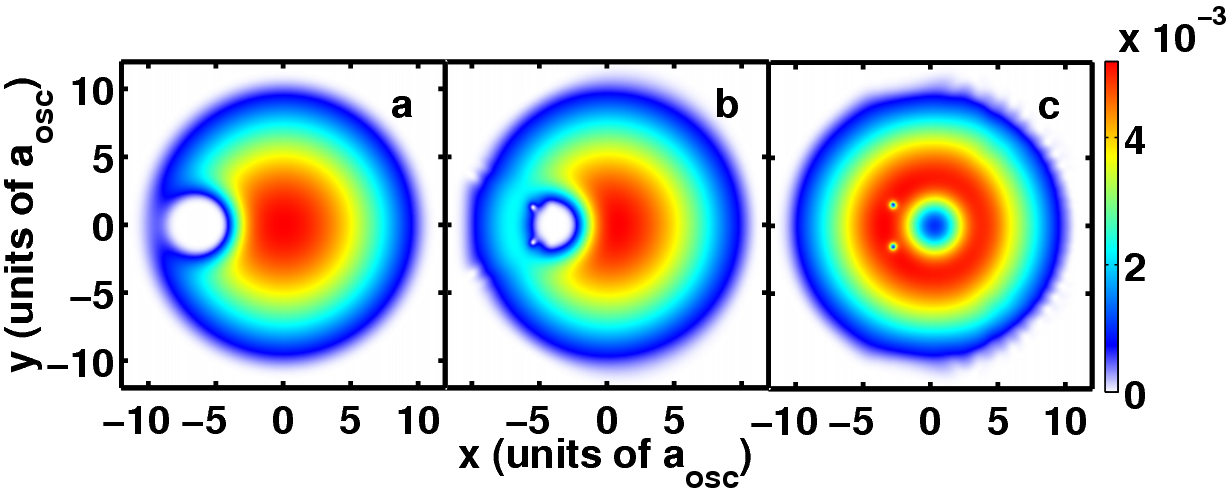}
 \caption{(Color online) Evolution of the superfluid Fermi gas in BCS limit
          when the obstacle is moving with the supercritical speed of 
         $175~\mu$m/s. Images with labels (a), (b), and (c) show the superfluid
         densities $|\phi(\mathbf r,t)|^2$ in units of $a_{\rm osc}^{-3}$ 
         on $xy$ plane at $t=0$ ms, $t=48$ ms, and $t=127$ ms respectively.
         }
\label{fig2}
\end{figure}

\begin{figure}[ht]
\begin{spacing}{0}% adjust the vertical spacing
\begin{tabular}{c}
%trim option's parameter order: left bottom right top
\resizebox{!}{!}
{\includegraphics[trim = 00mm 0mm 0mm 0mm,clip, angle=0,width=4.2cm]
                 {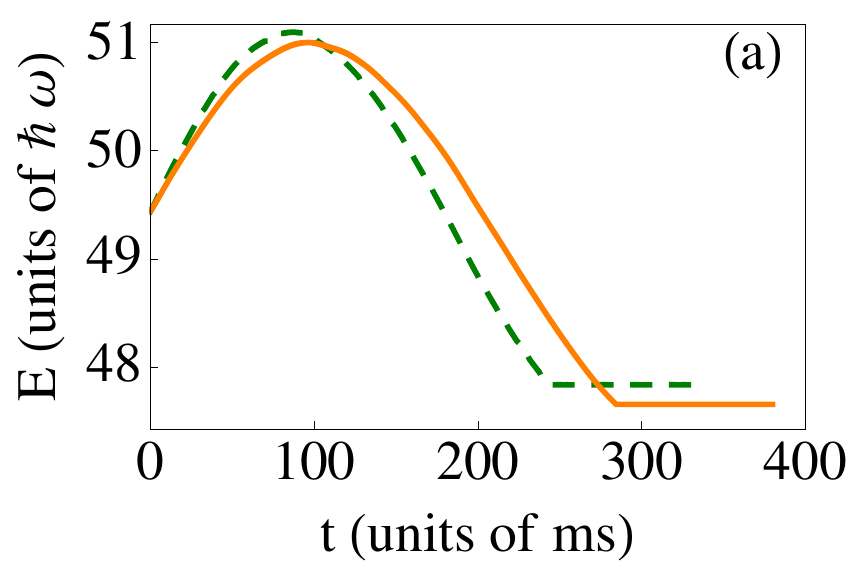}}
\resizebox{!}{!}
{\includegraphics[trim = 0mm 0mm 0mm 0mm,clip, angle=0,width=4.2cm]
                 {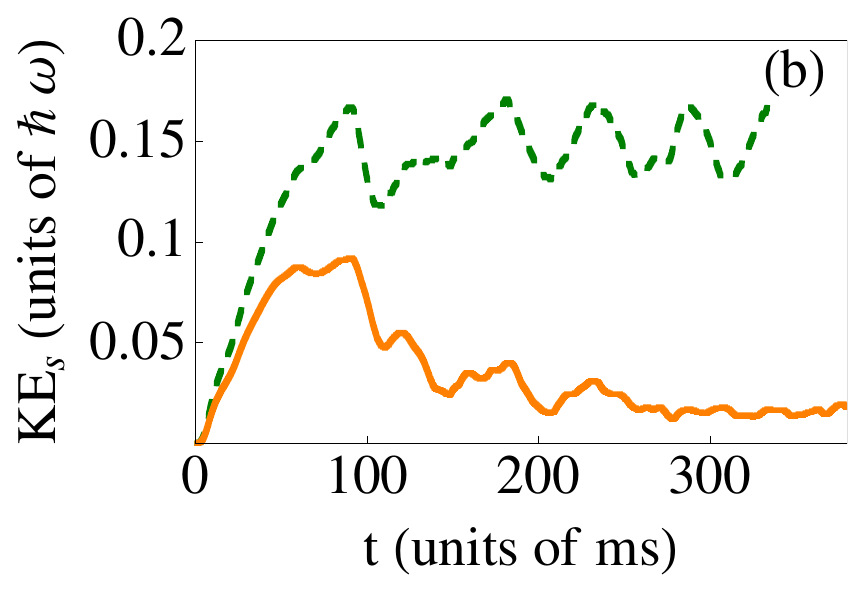}}\\
\end{tabular}
 \caption{(Color online) Variation of (a) $E$ and (b) $KE_s$ with time. 
         The dashed dark green curve shows the variation when 
         the obstacle is moving with the speed of $175~\mu$m/s. The solid 
         orange curve shows the variation when the obstacle
         is moving with the speed of $150~\mu$m/s.
         }
\label{fig11}
\end{spacing}
\end{figure}

We consider the repulsive Gaussian obstacle potentials with 
$V_0 = 93\hbar\omega$, $x_0 = -6a_{\rm osc}$, $w_0(0) = 10.0 \mu$m and move it 
along $x$-axis. While moving the obstacle we continuously decrease the 
strength of obstacle potential such that it vanishes at $x_0 = 6a_{\rm osc}$.
As is expected, we find that there is no generation of the vortices if the
obstacle is moved below the critical speed. For example, Fig. \ref{fig1} shows
the evolution of the superfluid density when the obstacle is moved with the
sub-critical speed of $150 \mu$m/s. 
On the other hand, when the obstacle is moved with a speed greater than the 
critical speed, vortex dipoles are produced. The generation of a vortex dipole 
for the obstacle potential moving with a super-critical speed of $175\mu$m/s is
shown in Fig. \ref{fig2}. The critical speed for vortex nucleation is a 
fraction of the speed of the local speed of sound. In case homogeneous BCS 
superfluid, the speed of the sound  \cite{PhysRevA.73.013607}
$ c = v_{\rm F}/\sqrt{3} $,
where $v_{\rm F} = 2 (3\pi^2 N |\psi|^2)^{1/3}$ (in scaled units) is the Fermi
velocity. For the parameters considered in the present work, the maximum speed
of sound at the trap center is $\sim 1465\mu$m/s. This is only an approximate
value since the presence of the trapping potential changes the speed of the 
sound from its value for homogeneous case
\cite{PhysRevA.73.021603}. 
When the obstacle moves across the superfluid with the sub-critical speed, it
can still generate sound waves. The generation of the sound waves can be 
inferred from the variation of superfluid kinetic energy with time. The
total kinetic energy $KE$ of the superfluid in scaled units is
\begin{eqnarray}
KE & = &\int\frac{\left(\nabla |\phi|\right)^2}{2}d\mathbf r + 
     \int\frac{\left(|\phi|\nabla \theta\right)^2}{2}d\mathbf r,\nonumber\\
   & = & KE_q + KE_s
\end{eqnarray}
where $\theta$ is the phase of superfluid wavefunction $\phi$.
The first energy term $KE_q$ is the quantum pressure energy and the second $KE_s$
is the energy arising from the superfluid velocity. After the obstacle 
potential has become zero, the total energy almost becomes constant as
is shown in Fig. \ref{fig11}(a). At the end of the time evolution, non-zero
value of $KE_s$ is mainly due to the presence of swirls, i.e. due to the 
vortices, and sound propagation; Fig. \ref{fig11}(b) shows the change in 
$KE_s$ with time.
In case when the obstacle moves with the speed of $175\mu$m/s, the $KE_s\sim 0.15\hbar\omega$
can be mainly ascribed to the swirls. On the other hand, when the obstacle moves 
with the speed of $150\mu$m/s, swirls are only present at boundary of the
SFG due to the creation of ghost vortices (vortices in low density regions). 
These ghost vortices do not contribute to $KE_s$. Hence the non-zero value
of $KE_s$ can be mainly ascribed to remnant sound waves in the superfluid. 
Although somewhat indistinct, the density ripples created by the sound
are present in the central region of Fig. \ref{fig1}(c).

\begin{figure}[ht]
\begin{spacing}{0}% adjust the vertical spacing
\begin{tabular}{c}
%trim option's parameter order: left bottom right top
\resizebox{!}{!}
{\includegraphics[trim = 00mm 0mm 0mm 0mm,clip, angle=0,width=8cm]
                 {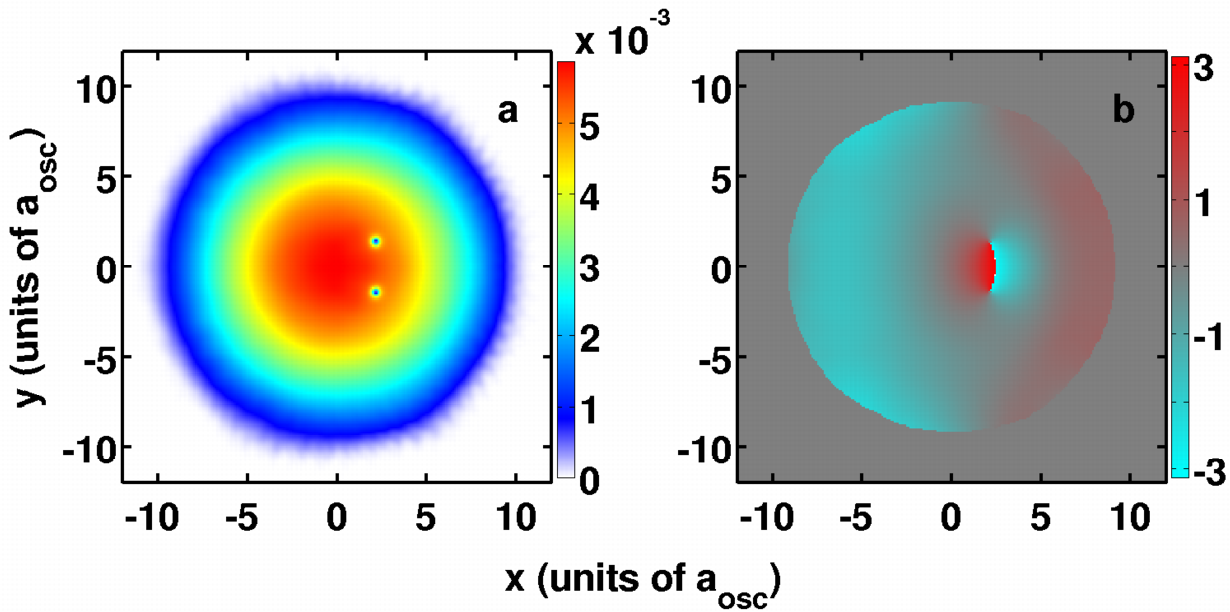}}\\
\resizebox{!}{!}
{\includegraphics[trim = 0mm 0mm 0mm 0mm,clip, angle=0,width=8cm]
                 {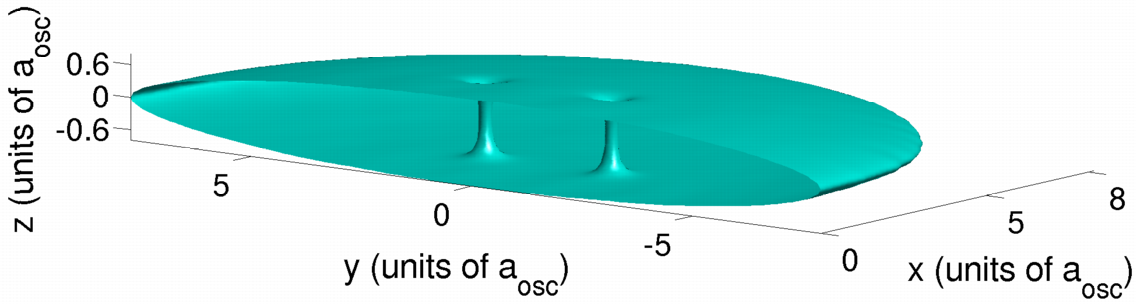}}\\
\end{tabular}
\caption{(Color online) In upper row, images with labels (a) and (b) show the 
         density and phase of the Fermi superfluid at $334$ ms. In lower row, 
         half of the isosurface of the superfluid at $334$ ms clearly shows 
         the presence of two straight vortex lines. The isosurface corresponds
         to $|\phi(\mathbf r)|^2 = 0.002a_{\rm osc}^{-3}$. 
}
\label{fig3}
\end{spacing}
\end{figure}

In the oblate condensates, the small length of vortex lines along axial 
direction (see Fig. \ref{fig3}) makes them less susceptible to vortex bending 
and hence vortex reconnections. This is due the suppression of Kelvon
generation in oblate superfluids \cite{PhysRevA.84.023637}. 
As a result, the vortex dipoles formed in 
pan-cake shaped traps are quite stable against reconnections and vortex 
antivortex annihilation. This is demonstrated in Fig. \ref{fig3}, here the
superfluid density, phase, and isodensity surface with isovalue of 
$0.002a_{\rm osc}^{-3}$ are shown after $334$ ms of evolution.  

We wish to emphasize that the NLSE can not account for the presence of normal
fluid in the system, and hence leads to the zero densities at the centers of
the vortices. Although very small at $T=0$ K, the normal fluid, which
predominantly occupies the vortex cores, significantly reduces the density 
depletion inside the vortex cores \cite{PhysRevA.64.063606,
PhysRevLett.87.100402,PhysRevLett.90.210402,PhysRevLett.91.190404,
PhysRevA.69.053622,PhysRevA.71.033631,PhysRevLett.96.090403}.

%%%%%%%%%%%%%%%%%%%%%%%%%%%%%%%%%%%%%%%%%%%%%%%%%%%%%%%%%%%%%%%%%%%%%%%%%%%%%%
%%%%        Shock waves and snake instability in BCS superfluids        %%%%%%
%%%%%%%%%%%%%%%%%%%%%%%%%%%%%%%%%%%%%%%%%%%%%%%%%%%%%%%%%%%%%%%%%%%%%%%%%%%%%%

\section{Shock waves and snake instability in BCS superfluids}
\label{III}
\begin{figure}[ht]
\begin{spacing}{0}% adjust the vertical spacing
\begin{tabular}{c}
%trim option's parameter order: left bottom right top
\resizebox{!}{!}
{\includegraphics[trim = 00mm 0mm 0mm 0mm,clip, angle=0,width=8cm]
                 {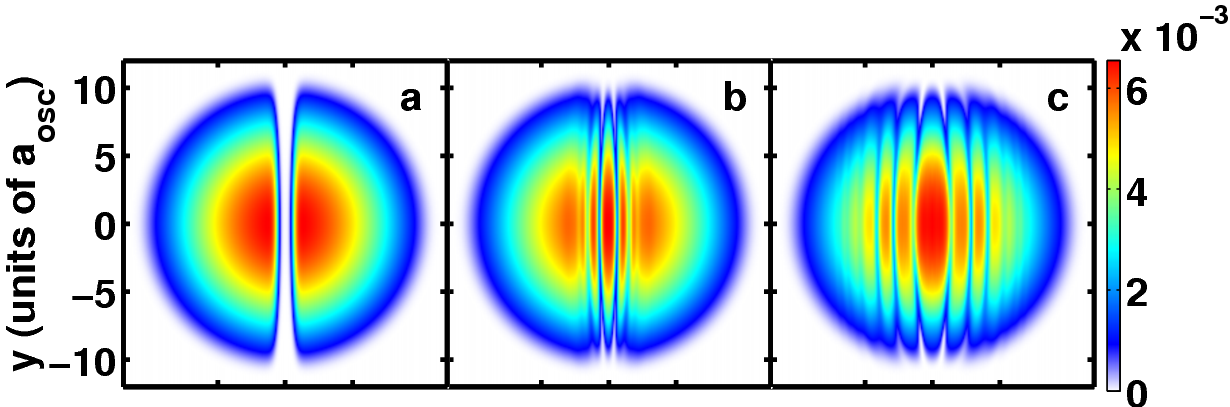}}\\
\resizebox{!}{!}
{\includegraphics[trim = 0mm 0mm 0mm 0mm,clip, angle=0,width=8cm]
                 {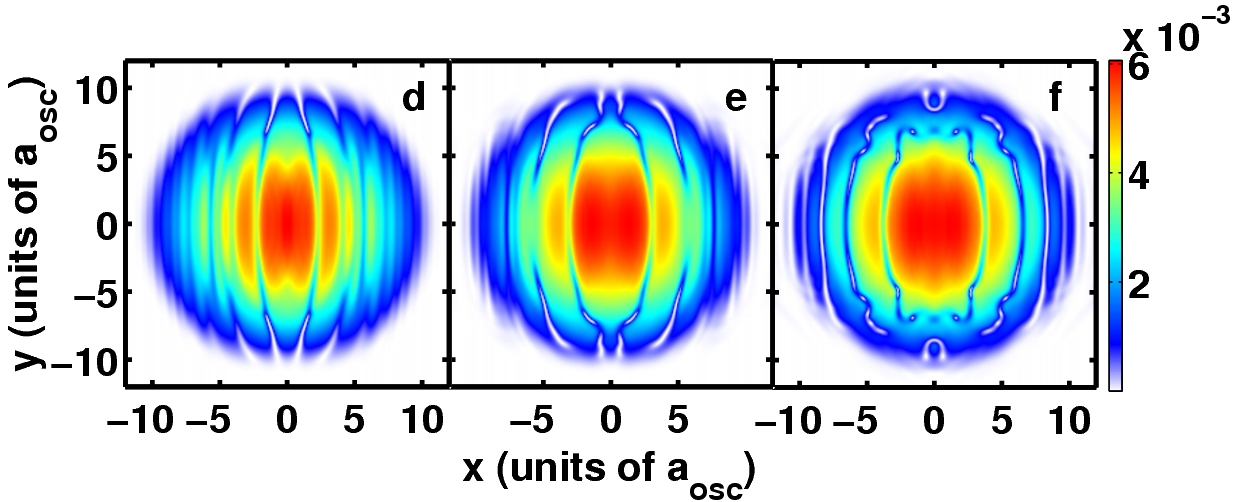}}\\
\end{tabular}
\caption{(Color online) Evolution of the superfluid density 
$|\phi(\mathbf r,t)|^2$ in units of $a_{\rm osc}^{-3}$ after the sudden 
merging of the two superfluid fragments. Images with labels (a), (b), (c),
(d), (e), and (f) are the densities at $t = 0$ ms, $4.8$ ms, $12.7$ ms,
$20.7$ ms, $28.6$ ms, and $36.6$ ms respectively. 
}
\label{fig4}
\end{spacing}
\end{figure}

\begin{figure}[ht]
\begin{spacing}{0}% adjust the vertical spacing
\begin{tabular}{c}
%trim option's parameter order: left bottom right top
\resizebox{!}{!}
{\includegraphics[trim = 00mm 0mm 0mm 0mm,clip, angle=0,width=8cm]
                 {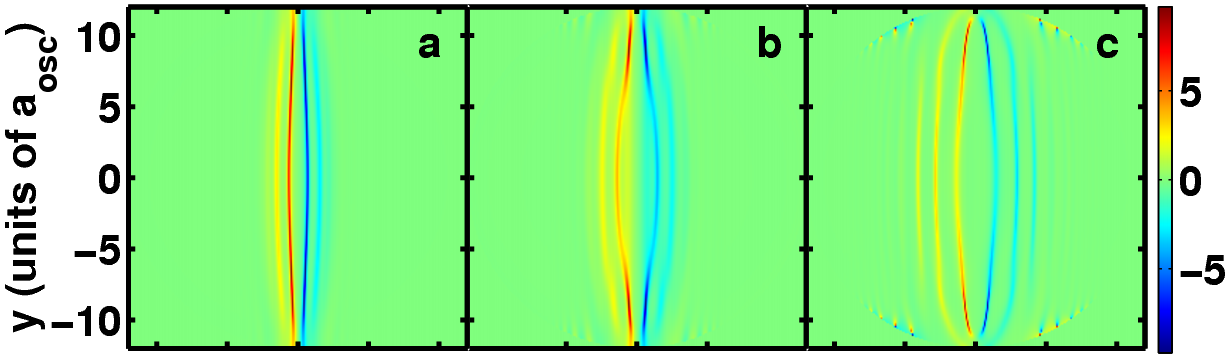}}\\
\resizebox{!}{!}
{\includegraphics[trim = 0mm 0mm 0mm 0mm,clip, angle=0,width=8cm]
                 {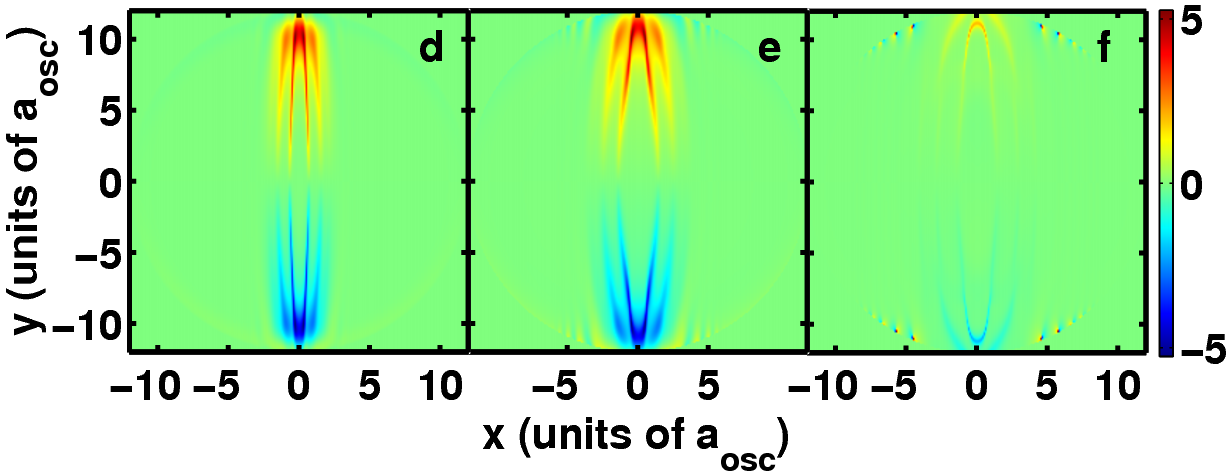}}\\
\end{tabular}
\caption{(Color online) The upper and lower row show the $x$ and $y$
components of the velocity field respectively, in scaled units
($a_{\rm osc}\omega$). 
Images with labels (a), 
(b) and (c) are the $x$ component of the velocity field at 
$t=4.8$ ms, $8$ ms, and $12.7$ ms respectively; exactly below them are the
corresponding $y$ component of the velocity field.
}
\label{fig5}
\end{spacing}
\end{figure}

As mentioned in the Sec. \ref{I}, soliton mediated decay of shock waves in
sufficiently wide traps can also lead to the formation of vortex dipoles.
To numerically study it, we create shock waves in the BCS superfluid Fermi gas 
by non-adiabatically merging the two superfluid fragments. The superfluid 
fragments are obtained by using the Gaussian obstacle potential along $y$-axis
and $z$-axis, 
i.e.,
\begin{equation}
V_{\rm obs}(\mathbf r,t) = V_0 \exp\left\lbrace -2\frac{\left[x-x_0(t)
                         \right]^2}{w_0^2}\right\rbrace.
\end{equation}
We consider $V_0 = 93~\hbar\omega$,  $x_0(t) = 0$, and $w_0 = 2.5~ \mu$m as the amplitude
and width of the obstacle potential. The stationary solution with this 
obstacle potential is shown in Fig. \ref{fig4}(a). 
We achieve the non-adiabatic merging of the two fragments of the superfluid 
by switching off the obstacle potential suddenly, and then letting the 
superfluid to evolve. After the obstacle potential is switched off at $t=0$ ms,
there is the formation central peak as is shown in Fig. \ref{fig4}(b) at 
$4.8$ms. This is consistent with the formation of central peak after the 
collision of two strongly interacting Fermi gas clouds in cigar-shaped trap 
\cite{joseph2011observation}. The central peak is bounded by two band solitons
which move outward as is shown in Fig. \ref{fig4}(c-d). The 
formation of the regions with large density gradients 
(see Fig. \ref{fig4}(b-c)) can be inferred as one of the signatures of the 
formation of shock wave. The development of large density gradients in 
superfluid, during the propagation of the shock wave, is also accompanied by 
large gradients in velocity field. Now, multiplying Eq. \ref{nlse} by 
$\psi^*$ and subtracting the resultant equation from its complex conjugate,
one gets
\begin{equation}
 \frac{\partial|\psi|^2}{\partial t} + \nabla.\left[\frac{\hbar}
 {4 m i}\left(\psi^*\nabla\psi-
   \psi\nabla\psi^*\right)
\right] =0 \nonumber
\end{equation}
Comparing it with the continuity equation for the superfluid density 
$|\psi(\mathbf r,t)|^2$, the velocity field of the superfluid is
\begin{equation}
 \mathbf v = \frac{\hbar}{4 m i}\frac{\left(\psi^*\nabla
  \psi - \psi\nabla\psi^*\right)}
 {|\psi|^2}.
\end{equation} 
Using this equation, the velocity field of the superfluid during the initial 
stages of the evolution is shown in Fig. \ref{fig5}, clearly showing the 
development of large velocity gradients. 
As is evident from Fig. \ref{fig5}, there are larger gradients in $x$ 
component of the velocity field as compared to $y$ component, indicative of 
the fact the potential barrier was along $y$ axis. 

\begin{figure}[ht]
\begin{spacing}{0}% adjust the vertical spacing
\begin{tabular}{c}
%trim option's parameter order: left bottom right top
\resizebox{!}{!}
{\includegraphics[trim = 00mm 0mm 0mm 0mm,clip, angle=0,width=8cm]
                 {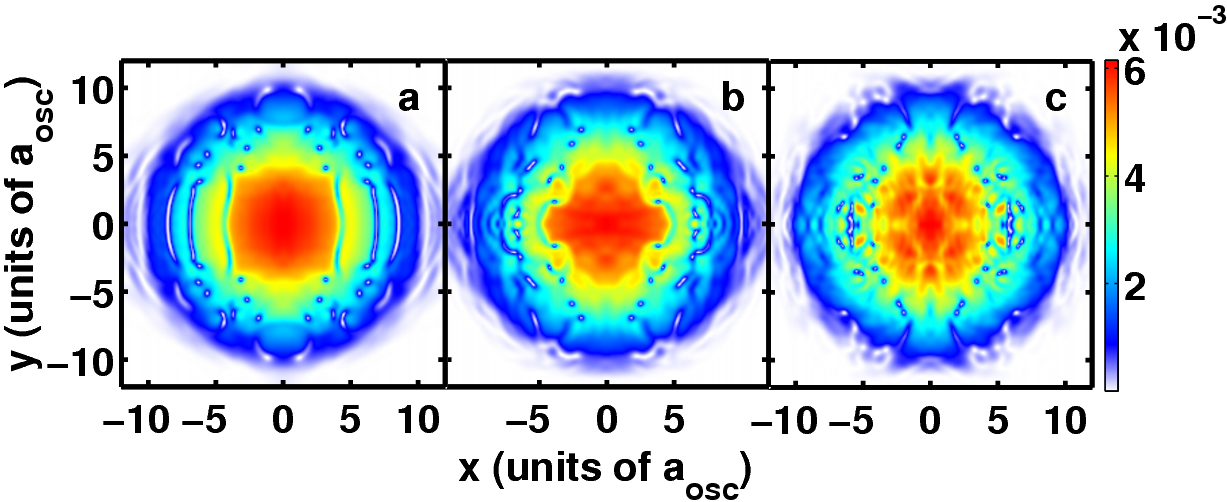}}\\
\resizebox{!}{!}
{\includegraphics[trim = 0mm 0mm 0mm 0mm,clip, angle=0,width=8cm]
                 {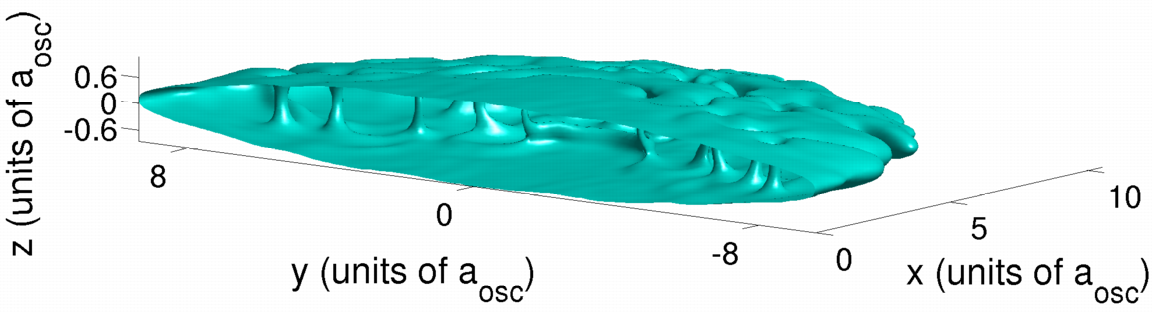}}\\
\end{tabular}
\caption{(Color online) Images with labels (a), (b), (c),
are the superfluid densities at $t=44.6$ ms, $52.5$ ms, and $60.5$ ms respectively.
In the lower panel, half of the isosurface of the superfluid at $t=52.5$ ms 
clearly shows the presence of five vortex dipoles. The isosurface corresponds
to $|\phi(\mathbf r)|^2 = 0.001a_{\rm osc}^{-3}$. 
}
\label{fig6}
\end{spacing}
\end{figure}

%\subsection{Decay of shock wave}
After some time, the soliton pair bounding the central peak starts to undergo 
snake instability [see Fig. \ref{fig4}(d-e)] leading to formation of the 
vortex dipoles [see Fig. \ref{fig4}(f) and \ref{fig6}(a)]. By this time, the outer
soliton pair also starts showing signatures of snake instability 
[see Fig. \ref{fig4}(f)]. This soliton pair also decays into vortex antivortex
pairs, Fig. \ref{fig6}(a-c). The three dimensional character of the vortex 
dipoles thus generated is evident from isodensity image in the lower
panel of Fig. \ref{fig6}.

\section{Conclusions}
\label{IV}
We have studied the generation of vortex dipoles in weakly interacting 
pancake-shaped Fermi superfluid at $T=0$K using a NLSE. We find that the 
supercritical motion of the repulsive obstacle across the BCS superfluid leads 
to the generation of vortex dipoles, whereas the sub-critical motion can 
generate only sound waves. We also observe the decay of shock waves, which 
are generated by sudden merging of the superfluid fragments, into vortex 
dipoles. The strong trapping force along axial direction provides 
stability to vortex dipoles against vortex reconnections and vortex-antivortex
annihilation events. This is responsible for the formation the 
dipoles which can persist for several seconds in the superfluid. 
%%%%%%%%%%%%%%%%%%%%%%%%%%%%%%%%%%%%%%%%%%%%%%%%%%%%%%%%%%%%%%%%%%%%%%%%%%%%%%%
%%%%%%%%%%%%             Acknowledgements                          %%%%%%%%%%%%
%%%%%%%%%%%%%%%%%%%%%%%%%%%%%%%%%%%%%%%%%%%%%%%%%%%%%%%%%%%%%%%%%%%%%%%%%%%%%%%

\begin{acknowledgements}
We thank Arko Roy, S. Chattopadhyay, Vivek Vyas and Dilip Angom for very 
useful discussions. The numerical computations reported in the paper were 
done on the 3 TFLOPs cluster at PRL. 
\end{acknowledgements}

%%%%%%%%%%%%%%%%%%%%%%%%%%%%%%%%%%%%%%%%%%%%%%%%%%%%%%%%%%%%%%%%%%%%%%%%%%%%%%%
%%%%                        Bibliography                                  %%%%%
%%%%%%%%%%%%%%%%%%%%%%%%%%%%%%%%%%%%%%%%%%%%%%%%%%%%%%%%%%%%%%%%%%%%%%%%%%%%%%%
\bibliography{references.bib} 
\end{document}